\documentclass[10pt,a4paper]{article}                                                                                                                                                                                                                                                         

\usepackage{graphicx}
\usepackage[linesnumbered,tworuled,vlined]{algorithm2e}
\usepackage{amsbsy,amsfonts,fixmath,isomath,theorem}

\usepackage[tbtags]{amsmath}
\usepackage{amssymb}
\DeclareMathOperator*{\argmin}{\arg\!\min}

\usepackage{pgf,pgfarrows,pgfnodes,tikz,color,tkz-berge}
\usetikzlibrary{arrows,shapes,snakes,automata,backgrounds,petri,topaths,calc}

\newcommand{\vect}[1]{\ensuremath{\mathbold{#1} } }

\newcommand{\R}{{\mathbb R}}
\newcommand{\Z}{{\mathbb Z}}

\newcommand{\D}[2]{ \ensuremath{ \frac{\mathrm{d} #1 }{\mathrm{d} #2 } }}
\newcommand{\C}{{\mathbb C}}

\begin{document}

\title{Analysis of Reaction Network Systems Using Tropical Geometry}

\author{Satya Swarup Samal$^1$, Dima Grigoriev$^2$,  Holger Fr\"ohlich$^1$, \\ and Ovidiu Radulescu$^3$   \\
\small  $^1$ Algorithmic Bioinformatics, Bonn-Aachen International Center for IT,  Bonn, Germany,  \\
\small  $^2$ CNRS, Math\'ematiques, Universit\'e de Lille, Villeneuve d'Ascq,
 59655, France, \\
\small  $^3$ DIMNP UMR CNRS 5235, University of Montpellier, Montpellier, France.
 }

\maketitle

\centerline{\bf Abstract}
We discuss a novel analysis method for reaction network systems with polynomial or
rational rate functions. This method is based on computing tropical equilibrations
defined by the equality of at least two dominant monomials of opposite signs in the
differential equations of each dynamic variable.
In algebraic geometry, the tropical equilibration problem is tantamount to
finding tropical prevarieties, that are finite intersections of tropical hypersurfaces.
Tropical equilibrations with the same set of dominant monomials
define a branch or equivalence class. Minimal branches are particularly
interesting as they describe the simplest states
of the reaction network. We provide a method
to compute the number of minimal branches
and to find representative tropical equilibrations
for each branch.

\section{Introduction}
\label{intro}

Networks of chemical reactions are widely used in chemistry for
modeling catalysis, combustion, chemical reactors, or in biology
as models of signaling, metabolism, and gene regulation. Several
mathematical methods were developed for analysis of these models
such as the study of multiplicity of  steady state solutions and
detection of
 bifurcations by stoichiometry analysis, deficiency theorems,
reversibility, permanency, etc. \cite{feinberg1987chemical}.

All these methods focus on the number and the stability of the
steady states of chemical networks. Beyond steady states,
metastable states defined as regions of the phase space where the
system has slow dynamics are also important for understanding the
behavior of networks. For instance, low dimensional inertial or
invariant manifolds gather slow degrees of freedom of the system
and are important for model reduction. Invariant manifolds can
lose local stability, which allow the trajectories to perform
large, rapid phase space excursions before slowing down in a
different place on the same or on another invariant manifold.
 \cite{haller2010localized}.
 Biological networks have often been modelled as discrete dynamical systems,
 whose trajectories are sequences of states in a discrete space, see, for instance, the Boolean automata
of Ren\'e Thomas \cite{thomas1973boolean}. We think that metastable states of continuous models are good candidates for representing states in a discrete model. It is, therefore, important to know how metastable states are dynamically connected.

We showed elsewhere that
tropical geometry methods can be used to approximate
such invariant manifolds for
systems of polynomial differential
equations \cite{NGVR12sasb,Noel2013a,Radulescu2015}. The connection between original dynamics of a specific system \cite{tyson1991modeling} to such a method based on invariant manifold  is described in \cite{NGVR12sasb}. In a nutshell, the slowness of the dynamics on the invariant manifolds
follows from the compensation of dominant forces acting on the system, represented
as dominant monomials in the differential equations. We have called
the equality of dominant monomials
tropical equilibration \cite{Noel2013a,Radulescu2015}. Tropical equilibrations are different from steady states,
because in tropically equilibrated systems one has non-compensated weak forces that drive the system slowly, whereas at steady state, the net forces vanish.
Furthermore, invariant manifolds
can be roughly associated to metastable states because they
are regions of phase space where systems dynamics is relatively
slower.
In this paper, we introduce methods to compute tropical equilibrations and group them into
branches that cover the metastable states of the system.
These branches of tropical equilibrations
form a polyhedral complex. The zeroth homology group (or in other words, the number
of connected components) of this complex indicates the possible
transitions between the metastable states. Additionally, we explore the structure between the equilibration solutions using directed graphs (to present the inclusion relations among them) and undirected graphs (to present the connectivity among them). Furthermore, one of the biochemical applications of these equilibrations is constructing invariant manifolds and to that extent defining slow variables. We provide a way to identify such slow species and visualise them through heatmaps. Lastly, we benchmark our method against models obtained from the Biomodels database \cite{Novere2006} and discuss the findings.

\section{Definitions and Settings}
\label{definitions}

We consider biochemical networks described by mass action kinetics
 \begin{equation}
 \D{x_i}{t} = \sum_j k_j S_{ij}  \vect{x}^{\vect{\alpha_{j}}}, \, 1 \leq i \leq n,
 \label{massaction}
 \end{equation}
where $k_j >0$ are kinetic constants, $S_{ij}$ are the
entries of the stoichiometric matrix (uniformly bounded integers,
$|S_{ij}| < s$, $s$ is small),
$\vect{\alpha}_{j} = (\alpha_1^j, \ldots, \alpha_n^j)$ are multi-indices,
  and $\vect{x^{\vect{\alpha}_{j}}}  = x_1^{\alpha_1^j} \ldots x_n^{\alpha_n^j}$.
 We consider that  $\alpha_i^j$ are non-negative integers.
 At this point, we like to mention, there exist approaches to describe such a polynomial system in a graph theoretic way, i.e., by a weighted directed graph and a weighted bipartite graph and to study the number of positive solutions depending on the graph structure \cite{Gatermann2002275}. This approach uses decompositions of Newton polytopes to find that parts of the directed graph are related to the existence of positive steady state solutions.  In our paper, we do not use these graph theoretic considerations and investigate the different problem of tropical equilibrations.

In the case of slow/fast systems with polynomial dynamics such as
 \eqref{massaction}, the slow invariant manifold
is approximated by a system of polynomial equations for the fast species.
This crucial point allows us to find a connection with tropical geometry.
We introduce now the terminology of tropical geometry needed for the presentation
of our results, and refer to
\cite{maclagan2009introduction} for a comprehensive introduction to this field.

Let $f_1, f_2, \ldots, f_k$ be  multivariate polynomials, $f_i \in \C[x_1,x_2,\ldots,x_n]$,
representing all or a part of the polynomials in the right hand side of  \eqref{massaction}.

Let us now assume that the variables $x_i,\, i \in [1,n]$ are written as powers of a small positive
parameter $\epsilon$, namely $x_i = \bar{x}_i \varepsilon^{a_i}$, where $\bar{x}_i$ has order
zero
(there are $c_i,d_i$ not depending on $\epsilon$ such that $0<c_i < \bar{x}_i < d_i$). The orders $a_i$ indicate the
order of magnitude of $x_i$. Because $\epsilon$ was chosen small, $a_i$ are lower for larger
absolute values of $x_i$. Furthermore, the order of magnitude of monomials $\vect{x^{\vect{\alpha}}}$
is given by the dot product  $\mu = \langle\vect{\alpha},\vect{a} \rangle$, where $\vect{a} = (a_1,\ldots,a_n)$.
Again, smaller values of $\mu$ correspond to monomials with larger absolute values.
For each multivariate polynomial $f$, we define the tropical hypersurface $T(f)$ as the set of vectors
$\vect{a} \in \R^n$ such that the minimum of $\langle\vect{\alpha},\vect{a} \rangle$ over all monomials in $f$
is attained for at least two
monomials in $f$. In other words, $f$ has at least two dominating monomials.

A {\em tropical prevariety} is defined as the intersection of a finite number of tropical
hypersurfaces, namely $T(f_1,f_2,\ldots,f_k) = \cap_{i\in[1,k]} T(f_i)$.

A {\em tropical variety} is the intersection of all tropical hypersurfaces in the ideal
$I$ generated by the polynomials $f_1,f_2,\ldots,f_k$,
$T(I) =  \cap_{f \in I} T(f)$.
The tropical variety is within the tropical prevariety,
but the reciprocal property is not always true. There exist algorithms to compute such tropical varieties as in \cite{Jensen2008}.

For our purpose, we slightly modify the classical notion of tropical prevariety.
A {\em tropical equilibration} is defined as a vector  $\vect{a} \in \R^n$ such that
$\langle\vect{\alpha},\vect{a} \rangle$ attains its minimum at least twice for monomials of different signs, for
each polynomial in the system $f_1,f_2,\ldots,f_k$. Thus, tropical equilibrations are subsets of
the tropical prevariety. Our sign condition is needed because we are interested in approximating
real strictly positive solutions of polynomial systems (the sum of several dominant
monomials of the same sign have no real strictly positive roots).

\section{Branches of Tropical Equilibrations}
\label{branches}

For chemical reaction networks with multiple timescales, it is reasonable to consider
that kinetic parameters have different orders of magnitudes.

We, therefore, assume that parameters of the kinetic models
\eqref{massaction} can be written as
\begin{equation}
k_j = \bar k_j \varepsilon^{\gamma_j}.
\label{scaleparam}
\end{equation}
The exponents $\gamma_j$ are considered to be integer. For instance, the following
approximation produces integer exponents:
\begin{equation}
\gamma_j = \text{round}( \log(k_j) / \log(\varepsilon)),
\end{equation}
where  round stands for the closest integer (with half-integers rounded to even numbers).
Without rounding to the closest integer, changing the parameter $\epsilon$
should not introduce variations in the output of our method. Indeed, the
tropical prevariety is independent of the choice of $\epsilon$.

Of course, kinetic parameters are fixed. In contrast, the orders of the species vary
 in the concentration space and have to be calculated as solutions to the tropical equilibration problem.
To this aim, the network dynamics is first described by a rescaled ODE system
 \begin{equation}
 \D{\bar{x}_i}{t} = \sum_j \varepsilon^{\mu_j - a_i} \bar k_j S_{ij}   {\bar{\vect{x}}}^{\vect{\alpha_{j}}},
 \label{massactionrescaled}
 \end{equation}
where
\begin{equation}
\mu_j(\vect{a}) = \gamma_j +  \langle \vect{a},\vect{\alpha_j}\rangle,
\label{muj}
\end{equation}
and $\langle , \rangle $ stands for the dot product. 

The r.h.s.\ of each equation in

\eqref{massactionrescaled} is a sum of multivariate monomials in the concentrations.
The orders $\mu_j$ indicate how large are these monomials, in absolute value.
A monomial of order $\mu_j$ dominates another monomial of order
$\mu_{j'}$ if
 $\mu_j < \mu_{j'}$.

{\em The tropical equilibration problem} consists in finding a vector $\vect{a}$ such that
\begin{equation}
\min_{j,S_{ij}  >0} ( \gamma_j + \langle \vect{a},\vect{\alpha_j}\rangle ) =
\min_{j,S_{ij}  <0} ( \gamma_j + \langle \vect{a},\vect{\alpha_j}\rangle )
\label{eq:minplus}
\end{equation}

This system can be represented as a set of linear inequalities resulting into a convex polytope. The solutions of this system have a
geometrical interpretation.
Let us define the extended order vectors $\vect{a}^e = (1, \vect{a}) \in \R^{n+1}$ and
extended exponent vectors $\vect{\alpha}_{j}^e= (\gamma_{j}, \vect{\alpha}_{j}) \in \Z^{n+1}$.
Let us consider the equality $\mu_j = \mu_{j'}$. This represents the equation of a
$n$ dimensional hyperplane of $\R^{n+1}$, orthogonal to the vector $\vect{\alpha_j}^e - \vect{\alpha_{j'}}^e$:
\begin{equation}
\langle \vect{a}^e,\vect{\alpha_j}^e \rangle = \langle \vect{a}^e,\vect{\alpha_{j'}}^e \rangle,
\label{lines}
\end{equation}
where $\langle,\rangle$ is the dot product in $\R^{n+1}$. We will see in the next section that
the minimality condition on the exponents $\mu_j$ implies that the normal vectors
$\vect{\alpha_j}^e - \vect{\alpha_{j'}}^e$
are edges of the so-called Newton polytope \cite{Henk2004,sturmfels2002solving}.

For each equation $i$, let us define
\begin{equation}
M_i(\vect{a}) = \underset{j}{\argmin}
 (\mu_j(\vect{a}), S_{ij}  >0) = \underset{j}{\argmin} (\mu_j(\vect{a}),S_{ij}  <0),
 \label{Mi}
\end{equation}
 in other words
$M_i$ denote the set of monomials having the same minimal exponent $\mu_i$.

We call {\em tropically truncated system}, the system obtained by keeping only the dominating monomials in \eqref{massactionrescaled}, as follows:

 \begin{equation}
 \D{\bar{x}_i}{t} = \varepsilon^{\mu_i - a_i} (\sum_{j\in M_i (\vect{a})} \bar k_j  S_{ij}  {\bar{\vect{x}}}^{\vect{\alpha_{j}}}).
 \label{massactionrescaledtruncated}
 \end{equation}

The tropical truncated system is uniquely determined by the index sets
$M_i(\vect{a})$, therefore, by the
tropical equilibration
$\vect{a}$. Reciprocally, two tropical equilibrations can have the same index sets
$M_i(\vect{a})$ and truncated systems. We say that two tropical equilibrations $\vect{a}_1$, $\vect{a}_2$
are  equivalent iff $M_i(\vect{a}_1) = M_i(\vect{a}_2), \text{for all } i$. Equivalence classes
of tropical equilibrations are called {\em branches}. For each branch there exists a unique convex polytope, cf.  \eqref{eq:minplus}. The union of branches are subsets of tropical prevariety. It is a subset because we are interested in tropical equilibration of at least two monomials of different signs as expressed in \eqref{eq:minplus}. This sign condition is essential as we are interested to approximate positive real solutions of the polynomial system.

\paragraph{Minimal Branches}
A branch $B$ with an index set $M_i$  is {\em minimal} if $M'_i \subset M_i$ for all $i$ where $M'_i$ is the index set of a branch $ B'$
implies $B' = B$ or $B'=$ emptyset. In the terminology of convex polytopes, this means a branch $B$ with a convex polytope $P_i$  is {\em minimal} if $P_i \subset P'_i$ for all $i$ where $P'_i$ is the convex polytope for branch $B'$ implies $B' = B$ or $B'$ is empty. For each index $i$, relation \eqref{lines} defines a hyperplane, the tropical equilibration branches are on intersections of $k$ such hyperplanes where $k$ is number of polynomial equations representing the right hand side of \eqref{massaction}. Minimal branches are maximal (w.r.t. inclusion) polytopes in the tropical prevariety.

\paragraph{Connected Components of Minimal Branches}
Two minimal branches represented by index sets $M_i$ and $M_j$ are connected if there exists a branch with index set $M_k$ such that $M_i\subset M_k$ and $M_j\subset M_k$. In the terminology of convex polytopes, this amounts to checking the intersection between two convex polytopes $P_i$ and $P_j$ (corresponding to minimal branches $M_i$ and $M_j$) if whether $P_i\cap P_j$ is non void  for all $i\neq j$.

\section{Algorithm}

In this section, we describe an algorithm to compute the tropical equilibrations, test the equilibrations for the equivalence classes (i.e., {\em branches}) and compute the {\em minimal branches} (cf. section \ref{branches}).
\subsection{Newton Polytope and Edge Filtering}\label{edge-filtering}
Given the input polynomial in the form of \eqref{massactionrescaled}, for each equation and species $i$, we define a Newton polytope ${\mathcal N}_i$, that is the convex hull of the set of  points $\vect{\alpha}_{j}^e$ such that $S_{ij} \neq 0$ and also including
together with all the points the half-line emanating from these points in the positive $\epsilon$ direction.
This is the Newton polytope of the polynomial in right hand side of \eqref{massactionrescaled},
with the scaling parameter $\varepsilon$ considered as a new variable.

\begin{figure}[h!]
\begin{center}
\includegraphics[width=0.7\textwidth]{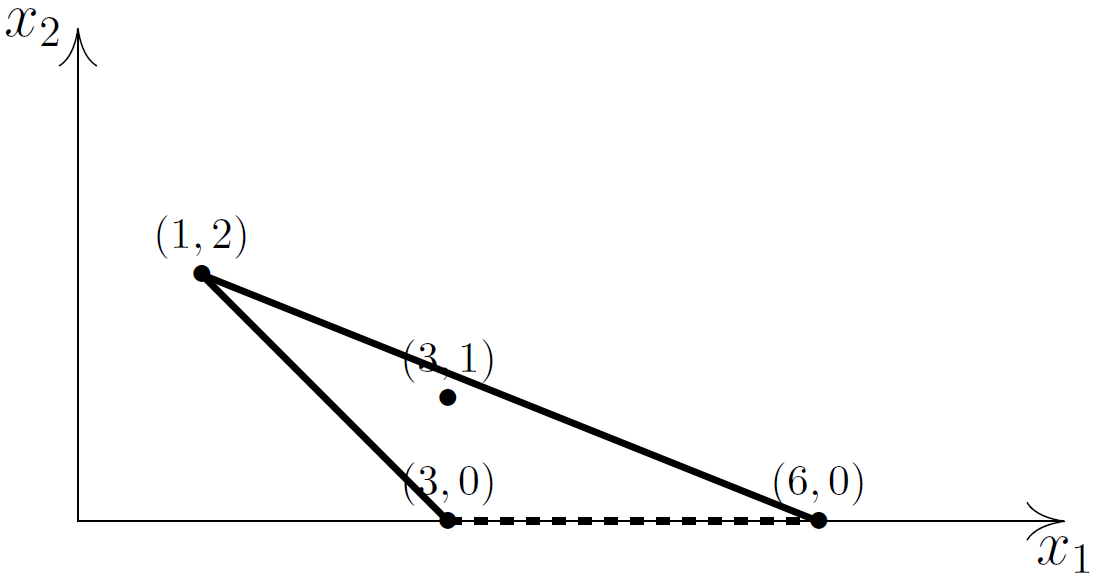}
\caption{ \label{fig:newton} \small
An example of a Newton polytope for the polynomial $-x_{1}^{6}+x_{1}^{3}x_{2}-x_{1}^{3}+x_{1}x_{2}^2$.
In this example, all the monomial coefficients have order zero in $\epsilon$ and we want to solve
the tropical problem $\min(3a_1+a_2, a_1 + 2a_2) = \min(6a_1,3a_1)$.
The Newton polytope vertices $(6,0),(3,0),(1,2)$ are connected by lines.
The point $(3,1)$ is not a vertex as it lies in the interior
of the polytope. This stems to having $\min(3a_1+a_2, a_1 + 2a_2)=a_1+2a_2$
for all tropical solutions, which reduces the number of cases to be tested.
The thick edges satisfy the sign condition, whereas the dashed
edge does not satisfy this condition.
For this example, the solutions of the tropical problem are in infinite
number and are carried by the two
half-lines $a_1=a_2 \geq 0$ and $5a_1= 2a_2 \leq 0$, orthogonal to the
thick edges of the Newton polygon
}
\end{center}
\end{figure}

As explained in section \ref{definitions}, the tropical equilibrations correspond to vectors $\vect{a}^e = (1,\vect{a}) \in \R^{n+1}$ satisfying the optimality condition as per \eqref{eq:minplus}. This condition is satisfied automatically on hyperplanes orthogonal to
edges of Newton polytope connecting vertices $\vect{\alpha}_{j'}^e$, $\vect{\alpha}_{j''}^e$
satisfying the opposite sign condition. Therefore, a subset of edges from the Newton polytope is selected based on the filtering criteria which tells that the vertices belonging to an edge should be from opposite sign monomials as explained in \eqref{eq:edge}.
\begin{eqnarray}
E(P)=\{\{v_{1},v_{2}\}\subseteq\left(_{2}^{V}\right)\mid \text{conv}(v_{1},v_{2})\in F_{1}(P) \nonumber \\
\wedge \, \text{sign}(v_{1})\times \text{sign}(v_{2})=-1\}, \label{eq:edge}
\end{eqnarray}
where $v_{i}$ is the vertex and $V$ is the vertex
set of the Newton polytope, $\text{conv}(v_{1},v_{2})$ is the convex hull of vertices
$v_{1},v_{2}$ and $F_{1}(P)$ is the set of 1-dimensional
face (edges) of the Newton polytope,
 {$\text{sign}(v_{i})$ represents the sign of the monomial which corresponds to vertex $v_{i}$. Figure \ref{fig:newton} shows an example of Newton polytope construction for a single equation.
Further definitions about properties of a polytope and Newton polytope can be found in
\cite{Henk2004,sturmfels2002solving}.}

\subsection{Computing Branches of Tropical Equilibrations} \label{computing-tropical}
\begin{figure}[h]
\begin{center}
\includegraphics[width=0.7\textwidth]{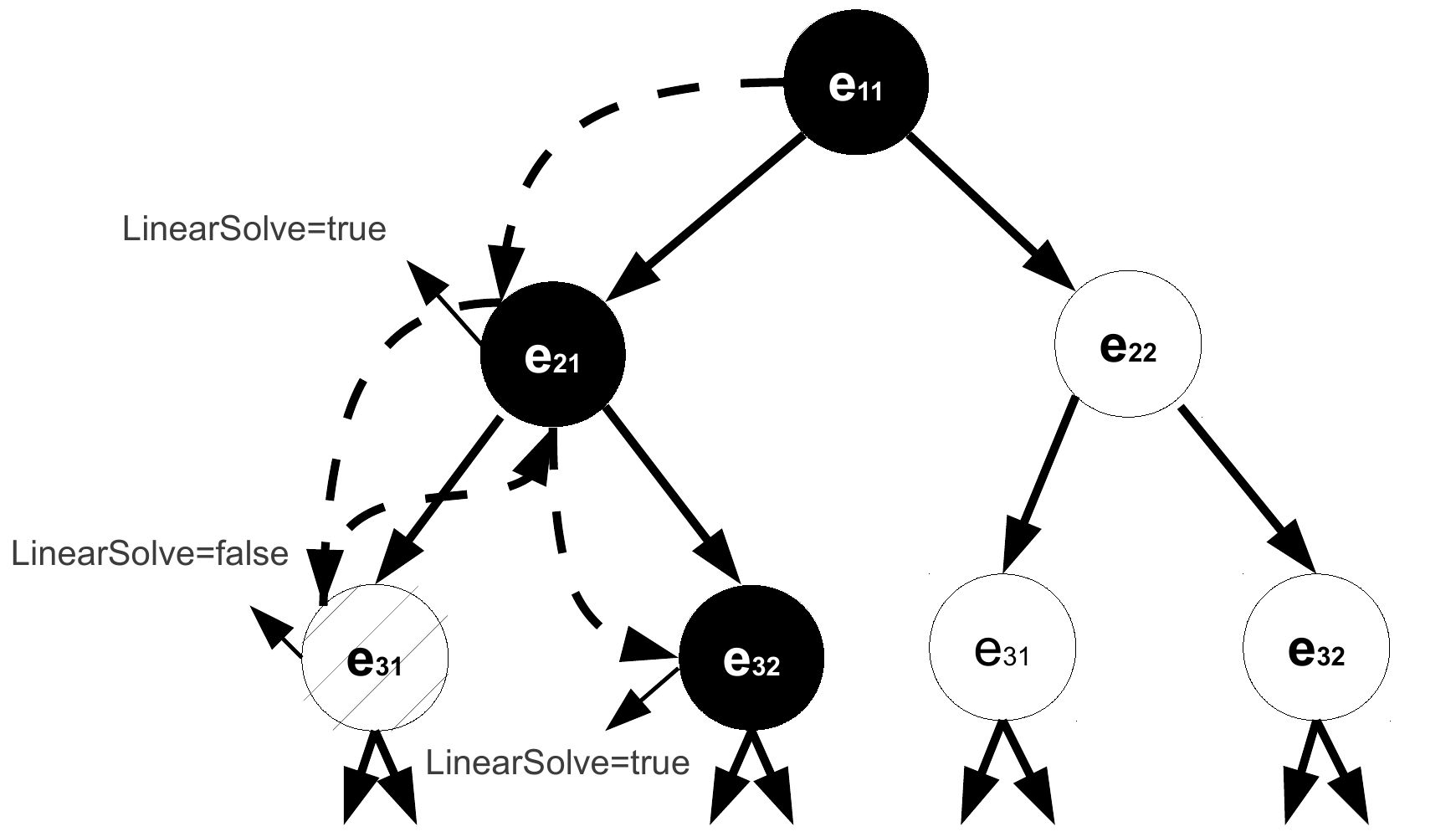}
\end{center}
\caption{\small \label{fig:pruning}Pruning strategy.
  {
The possible combinations of edges are represented in a tree representation where $e_{ij}$ represents $i$th edge from $j$th Newton polytope. An edge set  $ne_{i}$ is the set of edges for $i$th Newton polytope. The algorithm starts
by testing for feasible solution for first
pair of edge sets. If a feasible solution is found, the algorithm
proceeds further to other edge sets or it backtracks. In the figure,
$e_{11}$ and $e_{21}$ are selected from edge sets $ne_{1}$, $ne_{2}$
and are checked for a feasible solution satisfying \eqref{linprog}. If such a solution exists, it moves to $e_{31}$from the
next edge set and again checks for feasible solution, if not then
it backtracks to $e_{21}$ and then to $e_{32}$ which results in a feasible
solution. Therefore, the sub-tree with root node $e_{31}$ is discarded
from future searches and this improves running time. Likewise the branch $e_{11}$ and $e_{22}$ is explored. This approach is similar to the branch and bound algorithm technique.
The dashed arrows show the flow of the program}}
\end{figure}

Using the Newton polytope formulation, one can then solve the tropical equilibration problem in \eqref{eq:minplus} using the edges of Newton polytope (as in \eqref{Mi}). A feasible solution is a vector $(a_1,\ldots,a_n)$ satisfying all the equations of system \eqref{eq:minplus} and lies in the intersection of hyperplanes (or convex subsets of these hyperplanes) orthogonal to edges of Newton polytopes obeying the sign conditions. Of course, not all sequences of edges lead to non-void intersections and, thus, feasible solutions. This can be tested by the following
linear programming problem resulting from  \eqref{eq:minplus}:
\begin{equation}
\begin{split}
\gamma_j(i) + \langle \vect{a},\vect{\alpha_j(i)} \rangle =
\gamma_j'(i) + \langle \vect{a},\vect{\alpha_j'(i)} \rangle
\leq \gamma_j'' + \langle \vect{a},\vect{\alpha_j''} \rangle),\\ \text{for all} \, j'' \neq j,j', \nu_{j''i} \neq 0,\quad i =1,\ldots,n
\end{split}
\label{linprog}
\end{equation}
where $j(i),j'(i)$ define the chosen edge of the $i$th Newton
polytope. The set of indices $j''$ can be restricted to vertices
of the Newton polytope, because the inequalities are automatically
fulfilled for monomials that are internal to the Newton polytope.
From \eqref{linprog}, the sequence of edges leading to a feasible
solution is actually a set of linear inequalities and hence
constitutes a feasible solution system (convex polytope), which
was computed using Algorithm \ref{alg:Tropical-Equilibration}.
Such feasible solution systems are actually convex polytopes as
defined in \eqref{eq:minplus}. The equivalence classes among the
feasible solution systems constitute the equivalence classes of
tropical equilibrations called {\em branches}. This was done using
the method {\em equal\_polyhedra} implemented in the software
package polymake \cite{Gawrilow2000}. For instance, in the example
of the preceding section, the choice of the edge connecting
vertices $(1,2)$ and $(3,0)$ leads to the following linear
programming problem:
 $$a_1 + 2a_2 = 3a_1 \leq 6a_1, $$
 whose solution is a half-line orthogonal to the edge of the Newton polygon. The pseudo-code is presented in Algorithm \ref{alg:Tropical-Equilibration}. It is clear from the above that the possible choices are exponential. In order to improve the running time of the algorithm, the pruning strategy evaluates \eqref{linprog} in several steps(cf. Algorithm \ref{alg:Tropical-Equilibration} and Fig. \ref{fig:pruning}). It starts with an arbitrary pair of edges and proceeds to add the next edge only when the inequalities \eqref{linprog} restricted to these two pair of edges are satisfied. The pruning method is a heuristic to filter out the infeasible set of edge combinations. A similar approach was undertaken in \cite{Emiris1995117}.

\begin{algorithm}[htbp] 
 \DontPrintSemicolon
\caption{SolveOrders: Steps of tropical equilibration algorithm\label{alg:Tropical-Equilibration}}
\BlankLine
\KwIn
{ {List of  edge sets $ne_{1},ne_{2},...,ne_{n}$ (cf. Fig. \ref{fig:pruning}), and the corresponding vertices of Newton polytope}}
\KwOut{ Set of feasible solution systems (convex polytopes) corresponding to orders of the variables $\vect{a}_{1},\vect{a}_{2},...,\vect{a}_{n}$
(tropical equilibration solution set)}
\Begin{
solutionset =\{\}; integer $k$=1; equation = \{\}
\BlankLine
SolveOrders(equation, $k$, edge-sets, vertices)
\BlankLine
\If {$k>n$}{ return}
\For {$l=1$ to number of entries in $ne_{k}$edge-set}{
equation(k)* = vertices in $l^{th}$row
\BlankLine
inequalities* = all other vertices in $ne_{1}$to $ne_{k}$edge-sets
\BlankLine
\If{ LinearSolve(equation,inequalities)is feasible} {
\If{ $k=n$ } {add the feasible solution system to solutionset }
SolveOrders(equation, $k+1$,  $ne_{1},..,ne_{k}$, vertices)
}
}
}
*The equations and inequalities are initialised as per \eqref{linprog}
\end{algorithm}

\subsection{Computation of Minimal Branches} \label{minbranches}
\begin{figure}[h!]
\begin{tabular}{cc}
\includegraphics[width=0.5\textwidth]{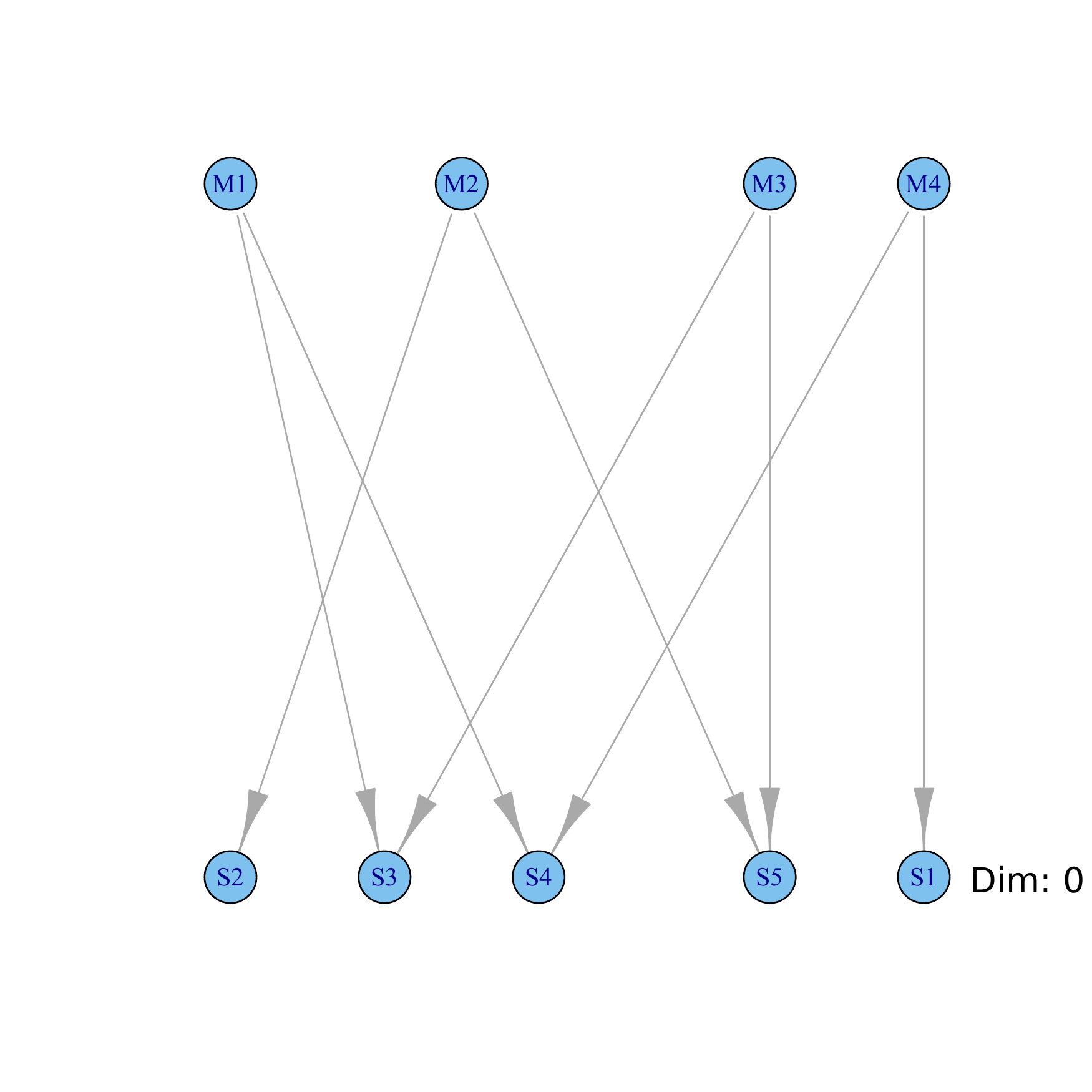} &
\includegraphics[width=0.5\textwidth]{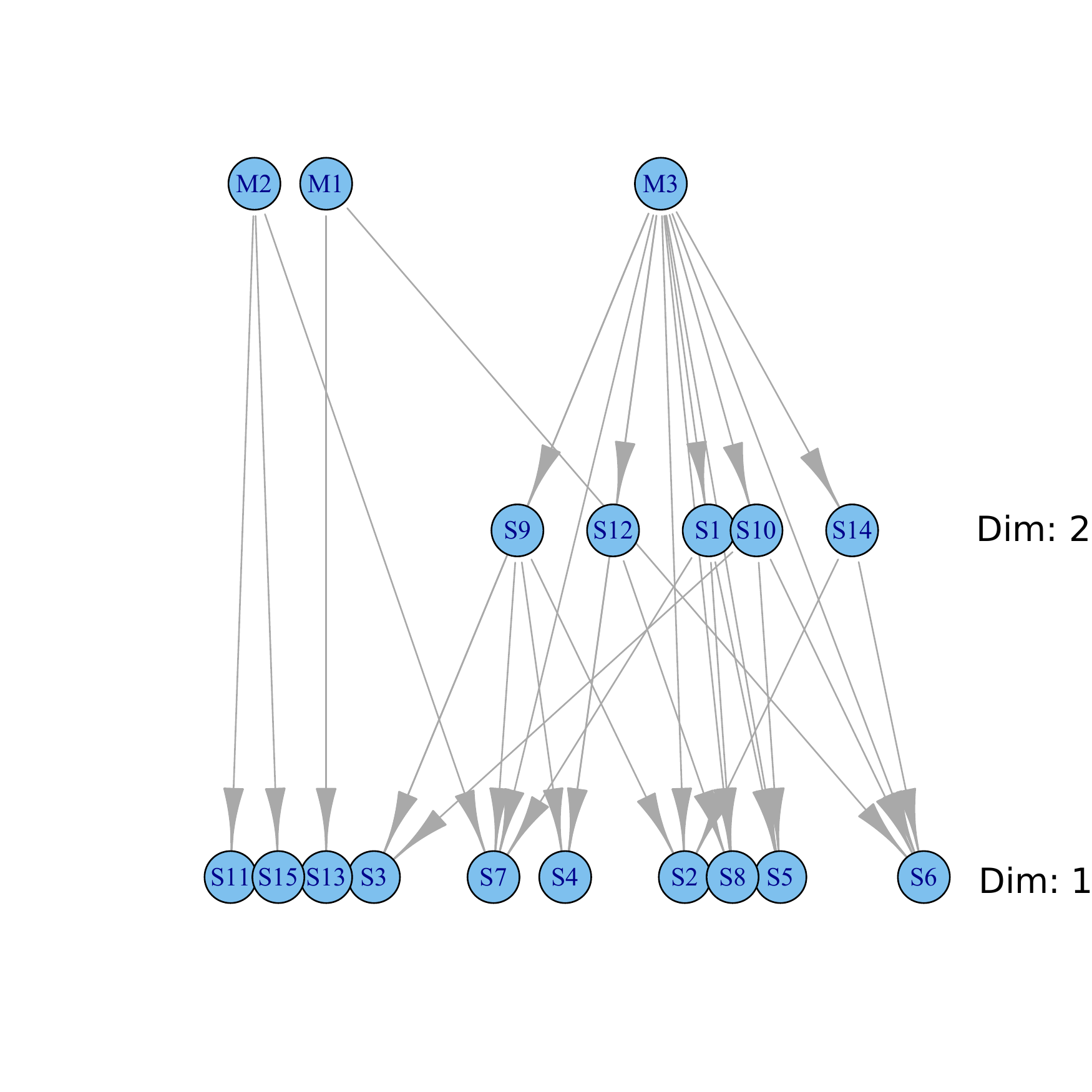} \\
{\footnotesize BIOMD0000000035 }& {\footnotesize BIOMD0000000072 }\\
\end{tabular}
\caption{ \label{fig:inclusion} \small
A directed graph in layered form showing the inclusion relations among the different solution branches (for $\varepsilon=1/11$) for two models namely BIOMD00000000-35,72. Vertices in the graph  comprise of polytopes corresponding to solution branches and an directed edge between $i$ and $j$ means $j$ is included in $i$. The topmost layer contain the minimal solution branches, thereafter the bottom layers are "included" solution branches. The layers of the included solution branches are based on the dimension of the corresponding polytopes (arranged in descending order). Therefore, included solutions in one layer are of same dimensional polytope  }
\end{figure}

The {\em minimal} branches are explained in section \ref{branches}. Computation was performed using {\em included\_polyhedra} method in polymake. The minimal branches as well as branches contained in minimal branches are represented in Fig. \ref{fig:inclusion} as a directed graph with layers where top most layer are minimal branches.

\subsubsection{Sample Point for Minimal Branches}
The polytopes corresponding to minimal solution branches obtained from the previous steps are represented by their facets and affine hull properties which are basically the set of inequalities and equalities. From such a set of inequalities, a sample point  $(a_1,\ldots,a_n)$ is computed using Satisfiability Modulo Theories (SMT) solver called Microsoft Z3 software \cite{DeMoura:2008:ZES:1792734.1792766} in python programming environment. With Microsoft Z3, one can generate the sample point belonging exclusively  to a minimal branch (and not at the intersections of minimal branches) by corresponding Boolean conjugations as shown in the following manner  \begin{equation}
\left\{ T_{i}\in B_{i} \wedge T_{i} \notin N_{i} ,\forall i\in I\right\} \label{eq:z3}
\end{equation}
where $T_{i}$ is a tropical equilibration solution corresponding to $B_i$ where $B$ is set of polytopes corresponding to minimal solutions  and $N$ is the set of rest solution branches along with minimal solution branches $ B\setminus B_i$ . $I$ is an index set denoting the elements of $B$.

The sample point thus obtained is a feasible solution to \eqref{linprog}. For our purpose, the benefit of using Z3 over any existing linear programming software is that it distinguishes strict and non-strict inequality conditions, which allows us to generate the sample point belonging exclusively  to a minimal branch.

\section{Results}
To compute the tropical equilibrations and to demonstrate the
running time of our algorithm, $33$ models  from the
r25 version of Biomodels database \cite{Novere2006} having
polynomial vector field were parsed with PoCaB \cite{Samal2012}.

\subsection{Summary}
A summary of the analysis is presented in Table \ref{tab:Summary-of-analysis}. The analysis is performed to compute all possible combinations of Newton polytope edges leading to tropical solutions within a maximal running time of $10000$ seconds of CPU time. In practice, we restrict this search space using the tree pruning strategy as explained in Fig. \ref{fig:pruning}. The analysis was repeated with four different choices for $\varepsilon$ values. In our framework, the number of variables may not be equal to the number of equations as the conservation laws (that are sums of variables whose total concentration is invariant) are treated as extra linear equations in our framework.

\subsection{Running Time}
\begin{figure}[h!]
\begin{center}
\includegraphics[width=0.94\textwidth]{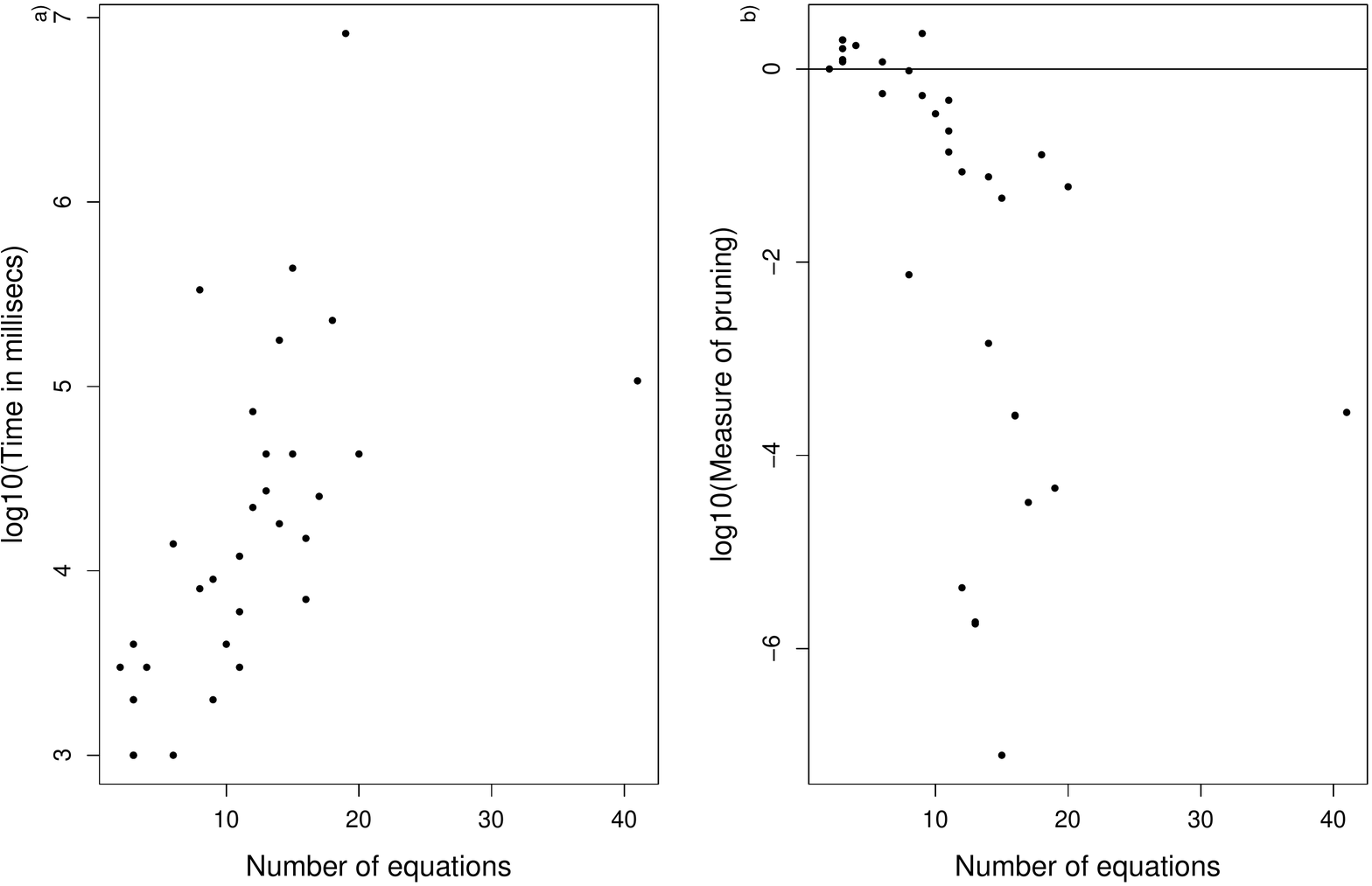}
\caption{ \label{fig:running-time} \small
(a) Plot of running time against number of equations in the model. (b) Pruning ratio for $\varepsilon$ value of $1/5$ against number of equations in the model}
\end{center}
\end{figure}

A semilog time-plot is presented in Fig. \ref{fig:running-time}(a) which plots the log of running time in milliseconds versus the number of equations in the model. In Fig. \ref{fig:running-time}(b),  the pruning ratio, i.e.,  the efficacy of tree pruning  for $\varepsilon$ value of $1/5$ is plotted. The pruning ratio is the ratio between the number of times the linear programming is invoked with every tree pruning step  (cf. Fig. \ref{fig:pruning}) and the possible number of combinations of Newton polytope edges possible without tree pruning. This ratio is, thus, a measure of efficiency achieved due to pruning.

\begin{table}[h!]
\caption{\small Summary of analysis on Biomodels database\label{tab:Summary-of-analysis}. Tropical solutions here mean existence of at least one feasible solution from all possible combination of vertices of the Newton polytope. Timed-out means all solutions could not be computed within $10000$ secs of computation time. No tropical solution implies that no possible combination of edges could be found resulting in a feasible solution. Model BIOMD0000000289 has solution at $\varepsilon$ values $1/5$,$1/7$ and $1/9$ but no solutions at $1/23$. Model BIOMD0000000108 has no solutions at all values of $\varepsilon$ }

\begin{small}
\begin{center}
\begin{tabular}{|c|r|r|r|r|r|r|}
\hline
\multicolumn{1}{|p{0.7cm}|}{$\varepsilon$ value} &
\multicolumn{1}{|p{1.2cm}|}{Total models con\-sidered }&
\multicolumn{1}{|p{1.5cm}|}{{Models without tropical solutions}} &
\multicolumn{1}{|p{1.5cm}|}{{Models with tropical solutions}} &
\multicolumn{1}{|p{1.5cm}|}{ Average running time (in secs)}&
\multicolumn{1}{|p{1.5cm}|}{ Average number of minimal branches}

\tabularnewline
\hline
\hline
1/5 & 33 & 1 & 32 & 299.38 & 3.24\tabularnewline
\hline
1/7 & 33  & 1 & 32 & 244.12 & 3\tabularnewline
\hline
1/9 & 33  & 1 & 32 & 309.73 & 3.75\tabularnewline
\hline
1/23 & 33  & 2 & 31 & 3179.32 & 3.84\tabularnewline
\hline
\end{tabular}
\end{center}
\end{small}
\end{table}

\subsection{Minimal Branches and Role of $\varepsilon$}
\begin{figure}[h]
\begin{center}
\includegraphics[width=0.94\textwidth]{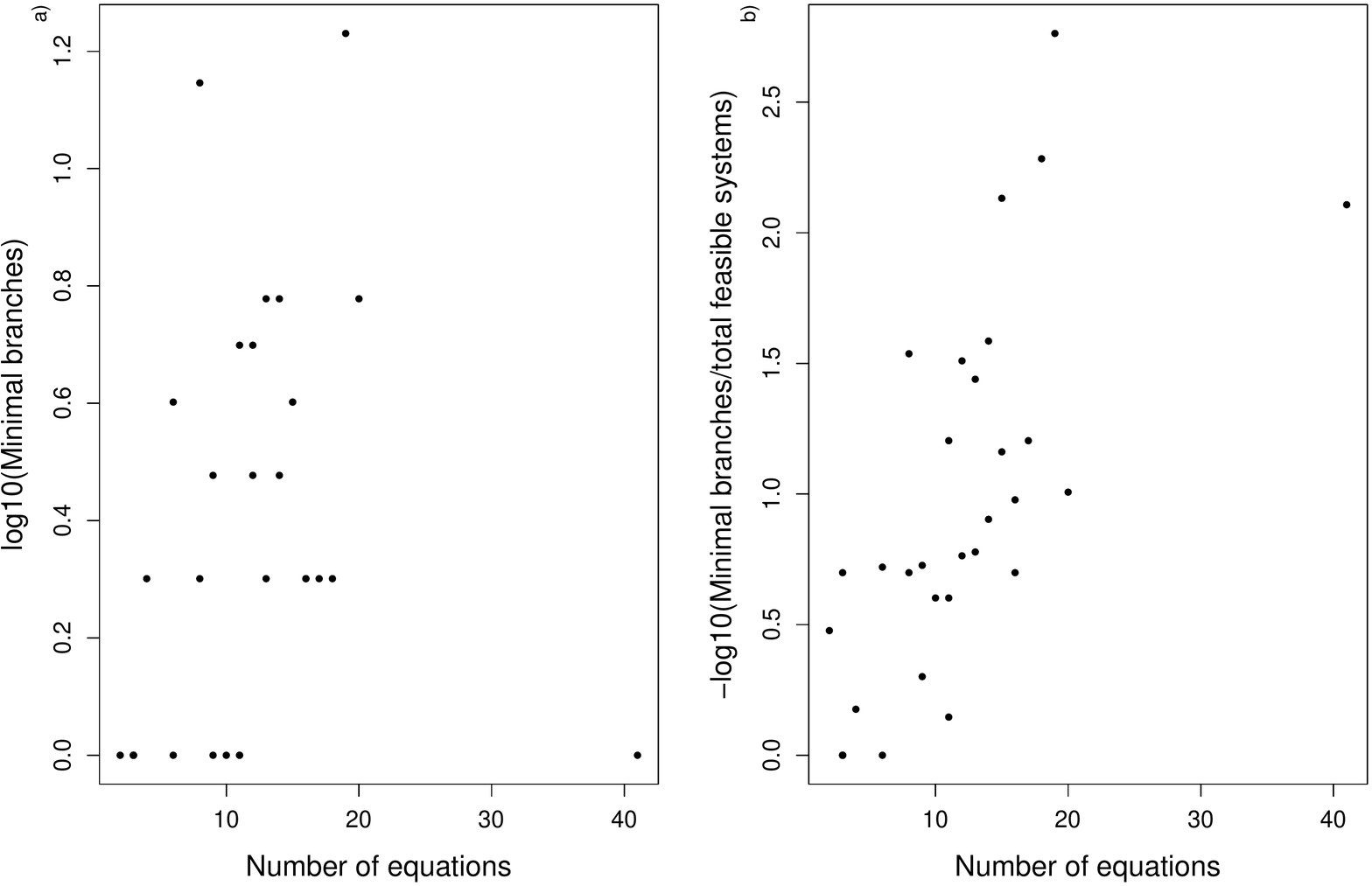}
\caption{ \label{fig:maximal} \small
(a) Minimal branches against number of equations in the model. (b) Ratio of minimal branches to the number of feasible solution systems, i.e., number of feasible edge combinations against number of equations in the model}
\end{center}
\end{figure}

A semilog plot for minimal solution branches is presented in Fig. \ref{fig:maximal}(a) and a semilog plot in Fig. \ref{fig:maximal}(b) showing the ratio of minimal solution branches to the number of feasible solution systems (obtained from Algorithm \ref{alg:Tropical-Equilibration}). It shows that a large proportion of feasible solution systems are either redundant or included in other feasible systems (i.e., inclusion relations).

\begin{figure}[h]
\begin{center}
\includegraphics[width=0.94\textwidth]{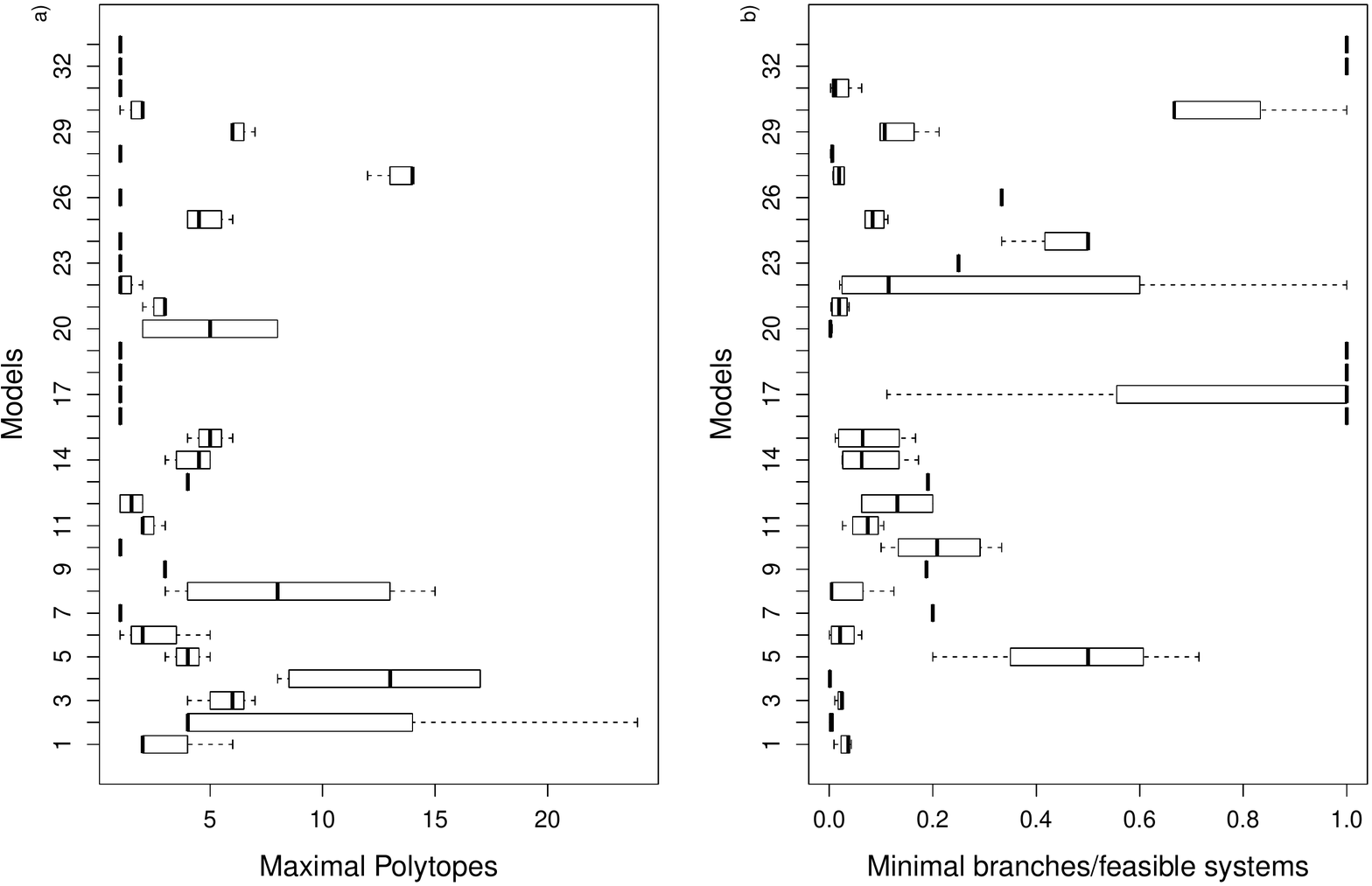}
\caption{ \label{fig:maximal-dist} \small
Boxplots showing (a) Distribution of Minimal branches. (b) Ratio of minimal branches to the number of feasible solution systems. Both distributions are at different $\varepsilon$ values: $1/5$,$1/7$,$1/9$,$1/23$}
\end{center}
\end{figure}

In order to investigate the effect of different  $\varepsilon$ values on the number of minimal solutions, a boxplot is presented in Fig. \ref{fig:maximal-dist}(a) for different choices of $\varepsilon$ values. In Fig. \ref{fig:maximal-dist}(b), the boxplot shows the ratio of minimal solution branches to the number of feasible solution systems (obtained from Algorithm \ref{alg:Tropical-Equilibration}) for different choices of $\varepsilon$ values

\subsection{Slow-Fast Variables}
\begin{figure}[h!]
\begin{tabular}{cc}
\includegraphics[width=0.5\textwidth]{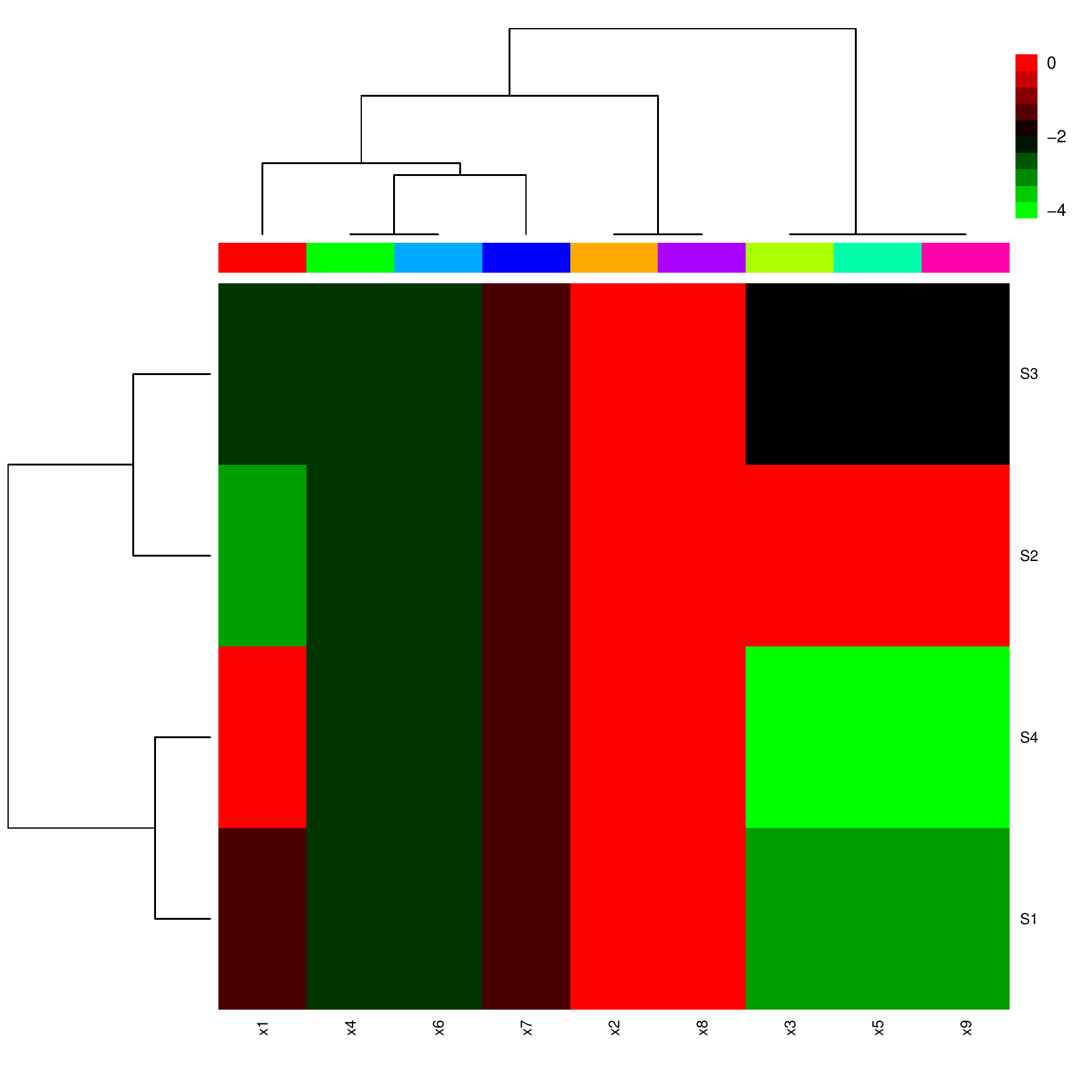} &
\includegraphics[width=0.5\textwidth]{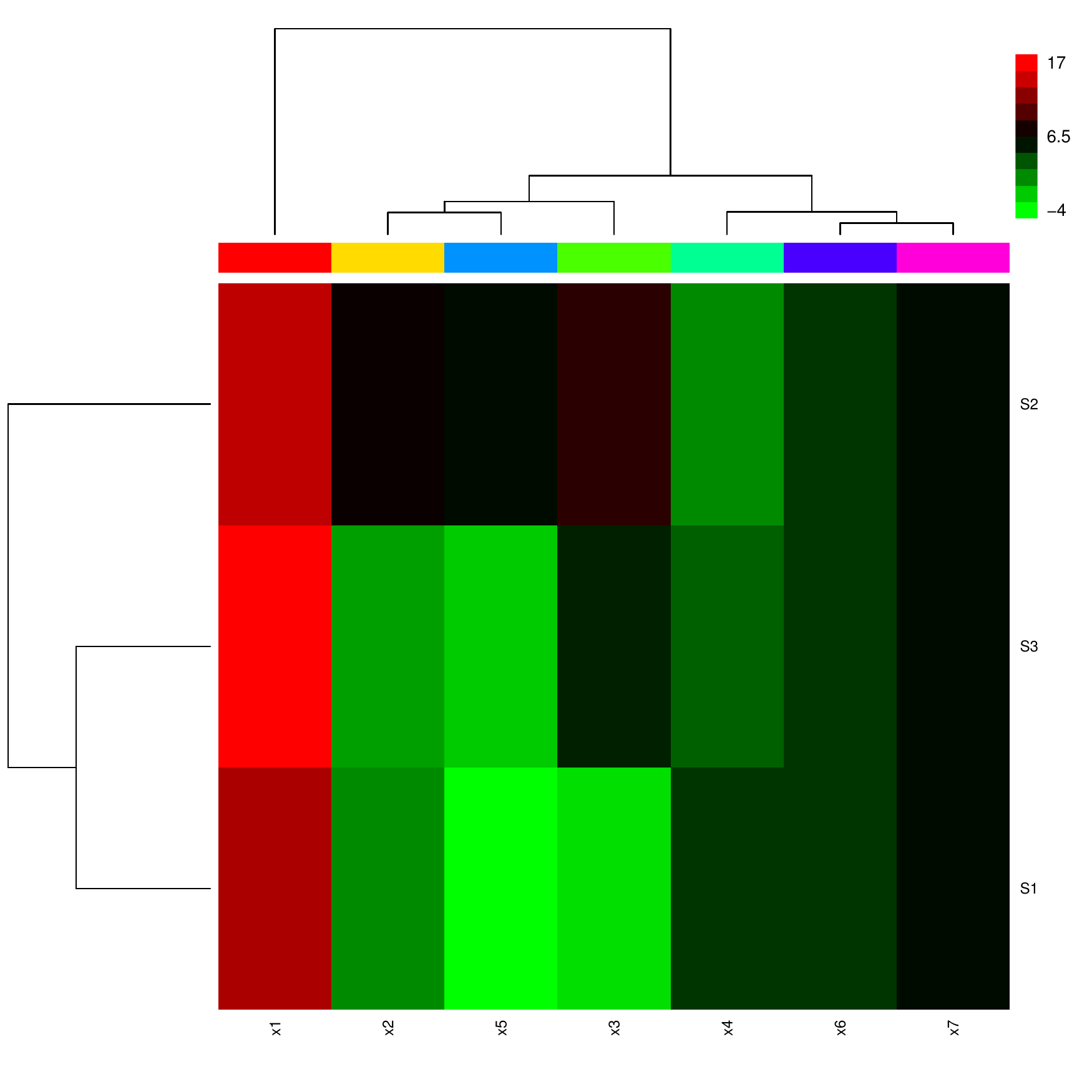} \\
{\footnotesize BIOMD0000000035 }& {\footnotesize BIOMD0000000072 }\\
\end{tabular}
\caption{ \label{fig:heatmap} \small
Heatmaps showing the rescaled orders (at $\varepsilon=1/11$) for four
models namely BIOMD00000000-35,72 with hierarchical clustering for
variables (horizontal axis) and tropical solutions (vertical axis). }
\end{figure}

\begin{figure}[h!]
\begin{tabular}{cc}
\includegraphics[width=0.5\textwidth]{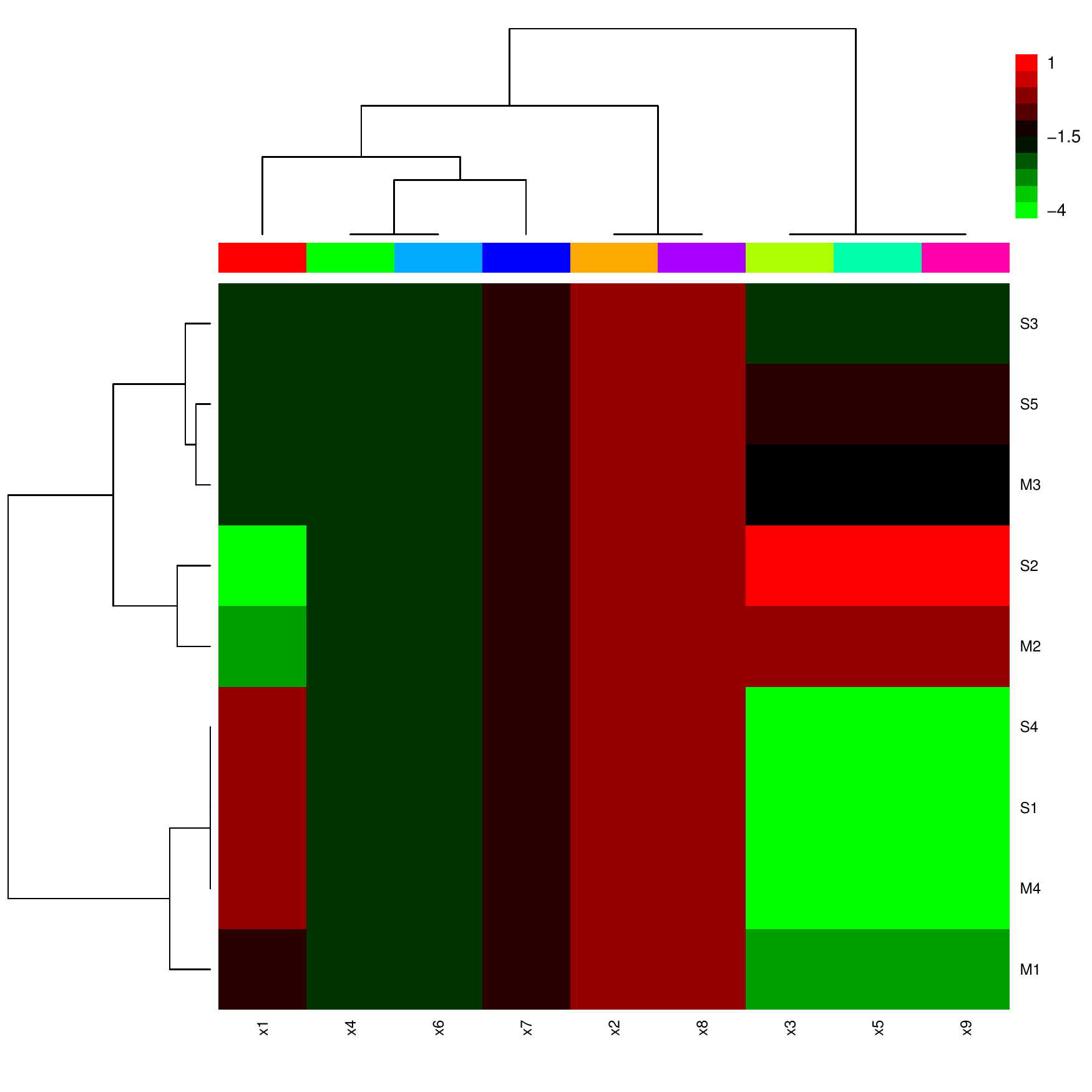} &
\includegraphics[width=0.5\textwidth]{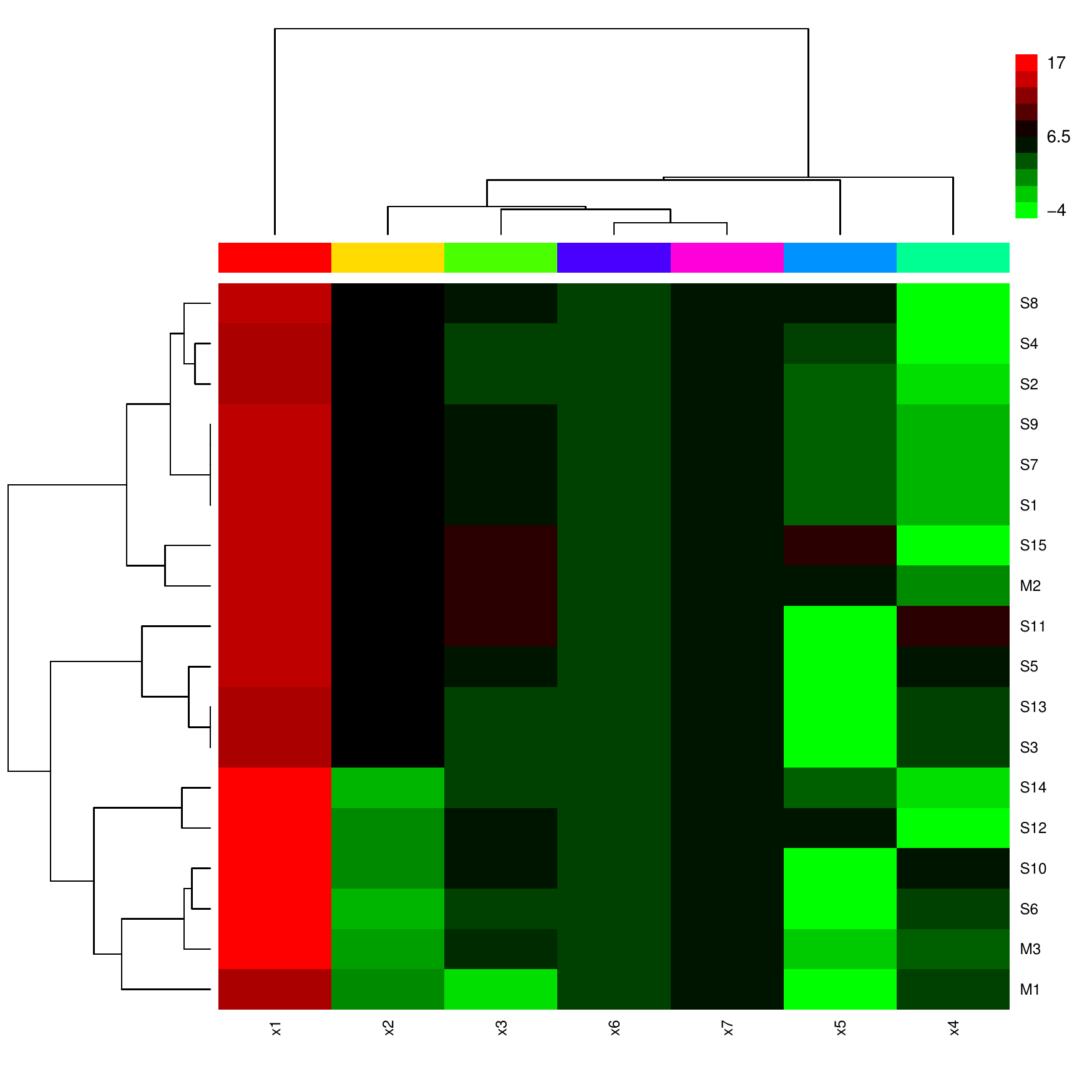} \\
{\footnotesize BIOMD0000000035 }& {\footnotesize BIOMD0000000072 }\\
\end{tabular}
\caption{ \label{fig:heatmap_all} \small
Heatmaps showing the rescaled orders (at $\varepsilon=1/11$) for four
models namely BIOMD00000000-35,72 with hierarchical clustering for
variables (horizontal axis) and tropical solutions (vertical axis). The heatmaps include the minimal solution branches (with prefix "M") and other solution branches excluding minimal ones (with prefix "S")}
\end{figure}

From the tropical equilibrations, we computed $\mu_j-a_i$(cf. \eqref{massactionrescaled}) which allow us to order the variables of the model, from the fastest (smallest $\mu_i - a_i$) to the slowest (largest $\mu_i - a_i$).  This is a measure to separate the variables into slow and fast which is an important step in constructing the invariant manifold for model reduction \cite{NGVR12sasb,Noel2013a,Radulescu2015,soliman2014constraint}. The heatmap in Fig. \ref{fig:heatmap} and Fig. \ref{fig:heatmap_all} plots this value for minimal solution branch and all solution branches, respectively, for few selected models. For some models, there appears to be a natural clustering (as seen from the dendogram from the hierarchical clustering) which requires further investigation.

\subsection{Connected Components}
\begin{figure}[h!]
\begin{tabular}{cc}
\includegraphics[width=0.5\textwidth]{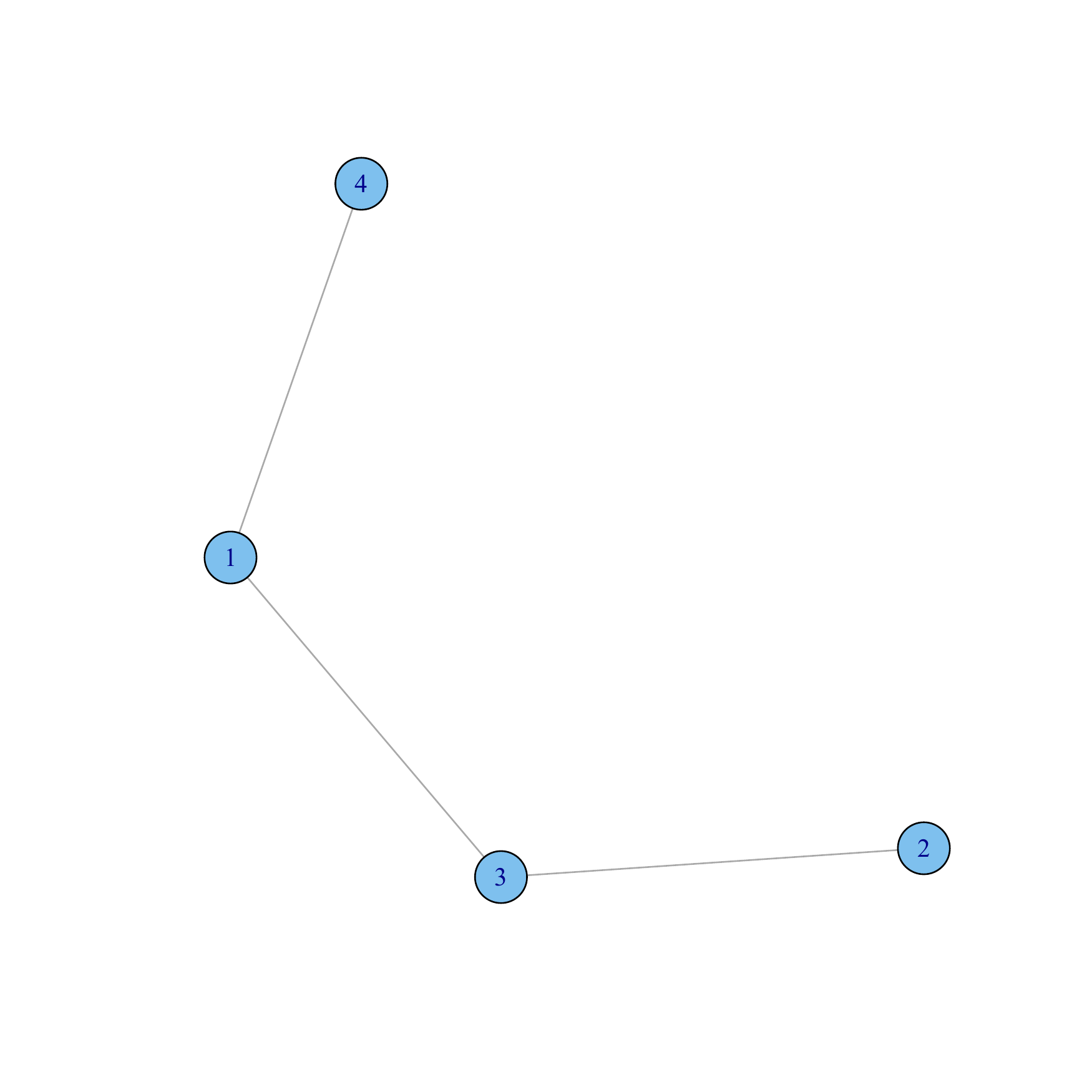} &
\includegraphics[width=0.5\textwidth]{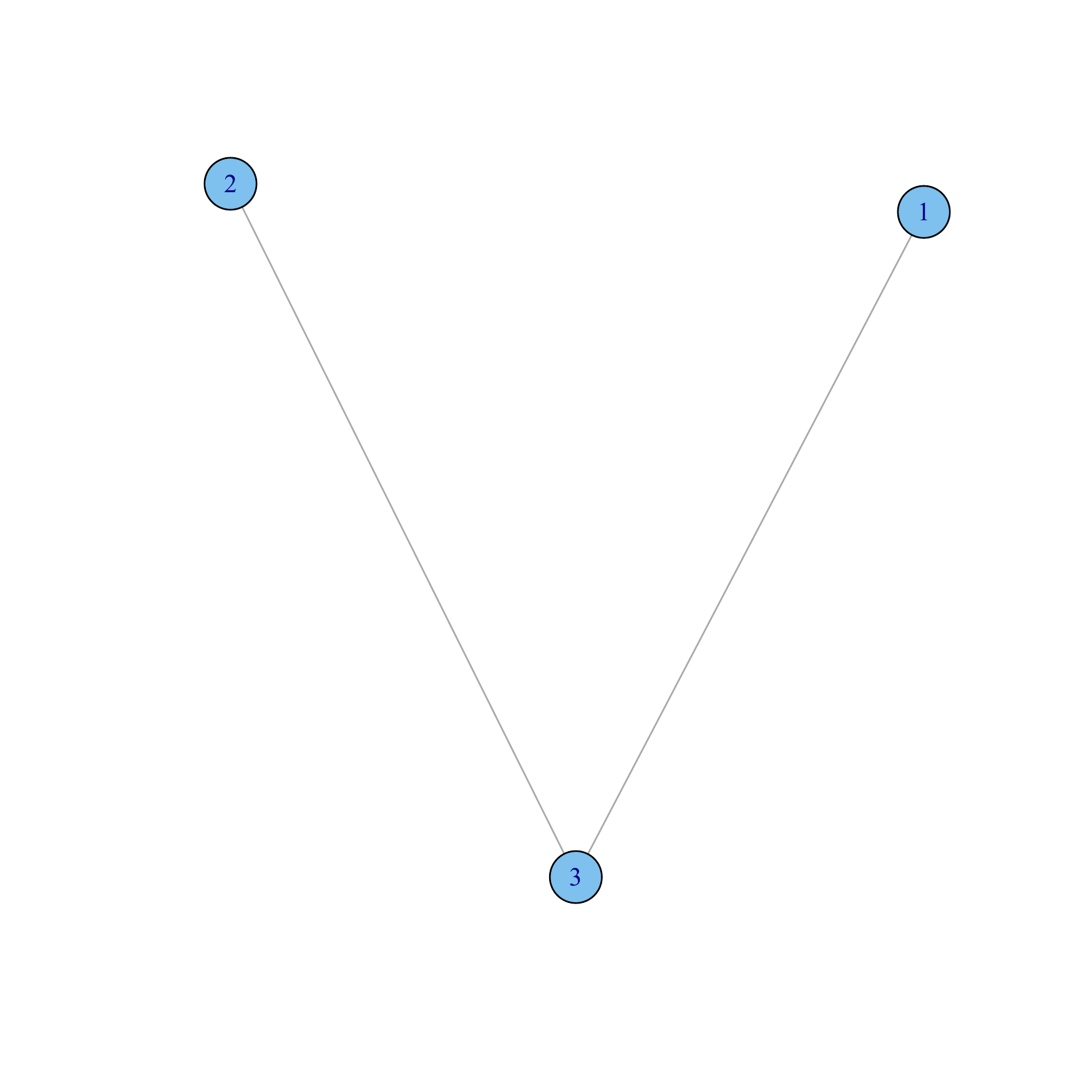} \\
{\footnotesize BIOMD0000000035 }& {\footnotesize BIOMD0000000072 }\\

\end{tabular}
\caption{\small
Graph of connected components (at $\varepsilon=1/11$)  for four models namely BIOMD00000000-35,72. All of them have one connected component. The vertices are minimal solution branch and there exists an edge if the intersection between the two vertices is non-void
\label{fig:connected}
}
\end{figure}

As described in section \ref{intro} and defined in section \ref{branches}, tropical solutions can be roughly associated with metastable states and these branches (which are convex polytopes) form a polyhedral complex. The number of connected components of this complex indicates the possible number of transitions between the minimal solution branches. Figure \ref{fig:connected} depicts such a graph of connected components of minimal solutions branches.

\section{Discussions}

We present an algorithm to compute the tropical equilibrations and organise them into {\em branches} and {\em minimal branches}. The directed graphs showing the inclusion relations between branches of tropical equilibration solutions reveal a rich structure. In addition, the connectivity of minimal branches are computed which provides an estimate for the possible dynamical transitions between them. One of the applications of tropical equilibration in systems biology is model reduction as demonstrated in \cite{Noel2013a,soliman2014constraint}. The concentration orders depicted in the heatmaps demonstrate the applicability of the algorithm in this direction. Thus, the overall solution structure provides insights into the dynamics of the system. More precisely, for the biochemical models in the Biomodels database, a large number of feasible solution systems were obtained but the number of minimal branches is considerably less.  For example, the number of minimal branches at $\varepsilon=1/5$ ranged from $1$ to $17$, whereas the total number of solutions ranged from $1$ to $9847$.

As the dominant terms in the polynomial system are the same for all the tropical solution
  on branches, it could be that branches correspond to invariant manifolds.
  In the same spirit, minimal branches could correspond to minimal invariant
  manifolds. This idea will be pursued in future work. Lastly, we have shown an application of tropical geometry to invariant manifolds defined by polynomial systems but the direct application of tropical geometry to differential equation systems \cite{Dima1991,Bruno:1086406} is also known.

\subsubsection*{Acknowledgements.}
This work has been supported by the French ModRedBio CNRS Peps, and EPIGENMED
Excellence Laboratory projects. D.G. is grateful to the
Max-Planck Institut f\"ur Mathematik, Bonn for its hospitality
during writing this paper and to  Labex CEMPI (ANR-11-LABX-0007-01). We are thankful to Prof. Dr. Andreas Weber (Institut fìr Informatik II, University of Bonn), Dr. Hassan Errami (Institut fìr Informatik II, University of Bonn) and Dr. Thomas Sturm (Max-Planck-Institut fìr Informatik, Saarbrìcken) for helpful discussions.


\end{document}